\documentclass[twocolumn,twoside,slac_two]{revtex4}
\usepackage{graphicx}
\usepackage{fancyhdr}
\begin{document}

\title{Testing GR with the Double Pulsar: Recent Results}


\author{M.~Kramer, D.R.~Lorimer, A.G.~Lyne, M.~McLaughlin}
\affiliation{University of Manchester, Jodrell Bank Observatory, UK}
\author{M.~Burgay, N.~D'Amico, A.~Possenti}
\affiliation{INAF - Osservatorio Astronomica di Cagliari, Italy}
\author{F.~Camilo}
\affiliation{Columbia Astrophysics Laboratory, Columbia University, USA}
\author{P.C.C. Freire}
\affiliation{NAIC, Arecibo Observatory, USA}
\author{B.C. Joshi}
\affiliation{NCRA, Pune, India}
\author{R.N.~Manchester, J. Reynolds, J.~Sarkissian}
\affiliation{Australia Telescope National Facility, CSIRO, Australia}
\author{I.H.~Stairs, R.~D.~Ferdman}
\affiliation{Department of Physics and Astronomy, University of
British Columbia, Canada}

\begin{abstract}
This first ever double pulsar system consists of two pulsars orbiting the
common center of mass in a slightly eccentric orbit of only 2.4-hr
duration. The pair of pulsars with pulse periods of 22 ms and 2.8 sec,
respectively, confirms the long-proposed recycling theory for
millisecond pulsars and provides an exciting opportunity to study the
works of pulsar magnetospheres by a very fortunate geometrical
alignment of the orbit relative to our line-of-sight. In particular,
this binary system represents a truly unique laboratory for
relativistic gravitational physics. This contribution serves as an
update on the currently obtained results and their consequences for
the test of general relativity in the strong-field regime. A
complete and more up-to-date report of the timing results will be
presented elsewhere shortly.
\end{abstract}

\maketitle

\thispagestyle{fancy}


\section{INTRODUCTION}

In this year 2005 we celebrate the work of Albert Einstein,
remembering his enormous contribution to our understanding of nature
and the Universe. One of the best ways for honoring his work is to
point out that still today, a hundred of years later, hundreds of
scientists around the world are deeply involved in searching for the
limits up to which his centennial theory of general relativity (GR)
can be applied. To date GR has passed all observational tests with
flying colours.  Still, it is the continued aim of many physicists
to achieve more stringent tests by either increasing
the precision of the tests or by testing different aspects. Some of
the most stringent tests are obtained by satellite experiments in the
solar system. One must not forget, however, that these solar-system
experiments are all made in the gravitational weak-field regime and
that they will never be able to provide tests in the strong-field
limit where deviations from GR may appear more clearly or even for the
first time (see e.g.~\cite{de98}). This strong-field regime is best
explored using radio pulsars.

Pulsars, highly magnetized rotating neutron stars,
 are unique and versatile objects which can be used to study an
extremely wide range of physical and astrophysical problems.  Beside
testing theories of gravity one can study the Galaxy and the
interstellar medium, stars, binary systems and their evolution, plasma
physics and solid state physics under extreme conditions.  In these
proceedings, we will present such applications for gravitational
physics made possible by the first ever discovered double pulsar
\cite{bdp+03,lbk+04}. We will demonstrate that this rare binary system
represents a truly unique laboratory for relativistic gravity. We will
present an update on the currently obtained timing results and their
consequences for tests of GR. A complete and more up-to-date report of
the timing results will be presented elsewhere shortly.

\section{THE DOUBLE PULSAR}

Our team discovered the 22.8-ms pulsar J0737$-$3039 in April 2003
\cite{bdp+03} in an extension to the hugely successful Parkes
Multi-beam survey \cite{mlc+01}. It was soon found to be a member of
the most extreme relativistic binary system ever discovered: its short
orbital period ($P_b = 2.4$ hrs) is combined with a remarkably
high value of periastron advance ($\dot{\omega}=16.9\deg$ yr$^{-1}$,
i.e.~four times larger than for PSR B1913+16!) that was
measurable after only a few days of observations. The system
parameters predict that the two members of the binary system will
coalesce on a short time scale of only $\sim 85$ Myr.  This
boosts the hopes for detecting a merger of two neutron stars with
first-generation ground-based gravitational wave detectors by a factor
of several compared to previous estimates based on only the
double neutron stars B1534+12 and B1913+16 \cite{bdp+03,kkl+04}.

In October 2003, we detected radio pulses from the second neutron star
when data sets covering the full orbital period were analysed
\cite{lbk+04}. The reason why signals from the 2.8-s pulsar companion
(now called PSR J0737$-$3039B, hereafter ``B'') to the millisecond
pulsar (now called PSR J0737$-$3039A, hereafter ``A'') had not been
found earlier, became clear when it was realized that B was only
bright for two short parts of the orbit.  For the remainder
of the orbit, the pulsar B is extremely weak and only detectable with
the most sensitive equipment. The detection of a young
companion B around an old millisecond pulsar A
confirms the evolution scenario proposed
for recycled pulsars (e.g.~\cite{bk74,sb76}) and made this already
exciting system sensational, providing a truly unique testbed for
relativistic gravity.

\section{STRONG-FIELD TESTS OF GENERAL RELATIVITY}

Since neutron stars are very compact massive objects, the double
pulsar (and other double neutron star systems)
can be considered as almost ideal point sources
for testing theories of gravity in the strong-gravitational-field
limit. Tests can be performed when a number of relativistic
corrections to the Keplerian description of an orbit, the so-called
``post-Keplerian'' (PK) parameters, can be measured. For point masses
with negligible spin contributions, the PK parameters in each theory
should only be functions of the a priori unknown neutron star masses
and the well measurable Keplerian parameters. 

With the two masses as the only free parameters, the measurement of
three or more PK parameters over-constrains the system, and thereby
provides a test ground for theories of gravity.  In a theory that
describes a binary system correctly, the PK parameters produce
theory-dependent lines in a mass-mass diagram that all intersect in a
single point.

As A has the faster pulse period (and is bright throughout the entire
orbit apart from a $\sim 27$-s
eclipse at superior conjunction), we can time A much
more accurately than B and measure precise PK parameters for A's
orbit. In GR, the five most important PK parameters are given 
to first post-Newtonian (1PN, or $\cal{O}$$(v^2/c^2)$) order by
\cite{dd86}:
\begin{widetext}
\begin{eqnarray}
\dot{\omega} &=& 3 T_\odot^{2/3} \; \left( \frac{P_b}{2\pi} \right)^{-5/3} \;
               \frac{1}{1-e^2} \; (M_A + M_B)^{2/3}, \label{omegadot}\\
\gamma  &=& T_\odot^{2/3}  \; \left( \frac{P_b}{2\pi} \right)^{1/3} \;
              e\frac{M_B(M_A+2M_B)}{(M_A+M_B)^{4/3}}, \\
\dot{P}_b &=& -\frac{192\pi}{5} T_\odot^{5/3} \; \left( \frac{P_b}{2\pi} \right)^{-5/3} \;
               \frac{\left(1 +\frac{73}{24}e^2 + \frac{37}{96}e^4 \right)}{(1-e^2)^{7/2}} \; 
               \frac{M_AM_B}{(M_A + M_B)^{1/3}}, \\
r &=& T_\odot M_B, \\
s &=& T_\odot^{-1/3} \; \left( \frac{P_b}{2\pi} \right)^{-2/3} \; x \;
              \frac{(M_A+M_B)^{2/3}}{M_B},
\end{eqnarray}
\end{widetext}
where $P_b$ is the period and $e$ the eccentricity and $x$ the
semi-major axis measured in light-s of the binary orbit. The masses
$M_A$ and $M_B$ of A and B, respectively, are expressed in solar
masses ($M_\odot$).  We define the constant
$T_\odot=GM_\odot/c^3=4.925490947 \mu$s where $G$ denotes the
Newtonian constant of gravity and $c$ the speed of light. The first PK
parameter, $\dot{\omega}$, is the easiest to measure and describes the
relativistic advance of periastron.  According to Eqn.~\ref{omegadot}
it provides an immediate measurement of the total mass of the system,
$(M_A+M_B)$. The parameter $\gamma$ denotes the amplitude of delays in
arrival times caused by the varying effects of the gravitational
redshift and time dilation (second order Doppler) as the pulsar moves
in its elliptical orbit at varying distances from the companion and
with varying speeds.  The decay of the orbit due to gravitational wave
damping is expressed by the change in orbital period, $\dot{P}_b$. The
other two parameters, $r$ and $s$, are related to the Shapiro delay
caused by the gravitational field of the companion. These parameters
are only measurable, depending on timing precision, if the orbit is
seen nearly edge-on. For pulsar A, all these quantities have indeed
been measured, providing a large number of available tests. In fact,
in addition to tests with these PK parameters, the possibility to
measure the orbit of both A and B opens up opportunities that go well
beyond what is possible with previously known double neutron stars, as
we will describe now.

With a measurement of the projected semi-major axes of the orbits of
both A and B, we obtain a precise measurement of
the mass ratio, $R(M_A, M_B)$, from Kepler's third law,
\begin{equation}
R(M_A,M_B) \equiv M_A/M_B = x_B/x_A.
\end{equation}
For every realistic theory of gravity, we can expect the mass ratio,
$R$, to follow this simple relation \cite{dt92}, at least to 1PN
order. Most importantly, the $R$ value is not only theory-independent,
but also independent of strong-field (self-field) effects which is not
the case for PK-parameters.  This provides a stringent and new
constraint for tests of gravitational theories as any combination of
masses derived from the PK-parameters {\em must} be consistent with
the mass ratio derived from Kepler's 3rd law. With five PK
parameters already available, this additional constraint makes the
double pulsar the most overdetermined system to date where the
most relativistic effects can be studied in the strong-field limit.

\section{TIMING OF THE DOUBLE PULSAR}

\begin{figure*}
\centering
\includegraphics[width=120mm]{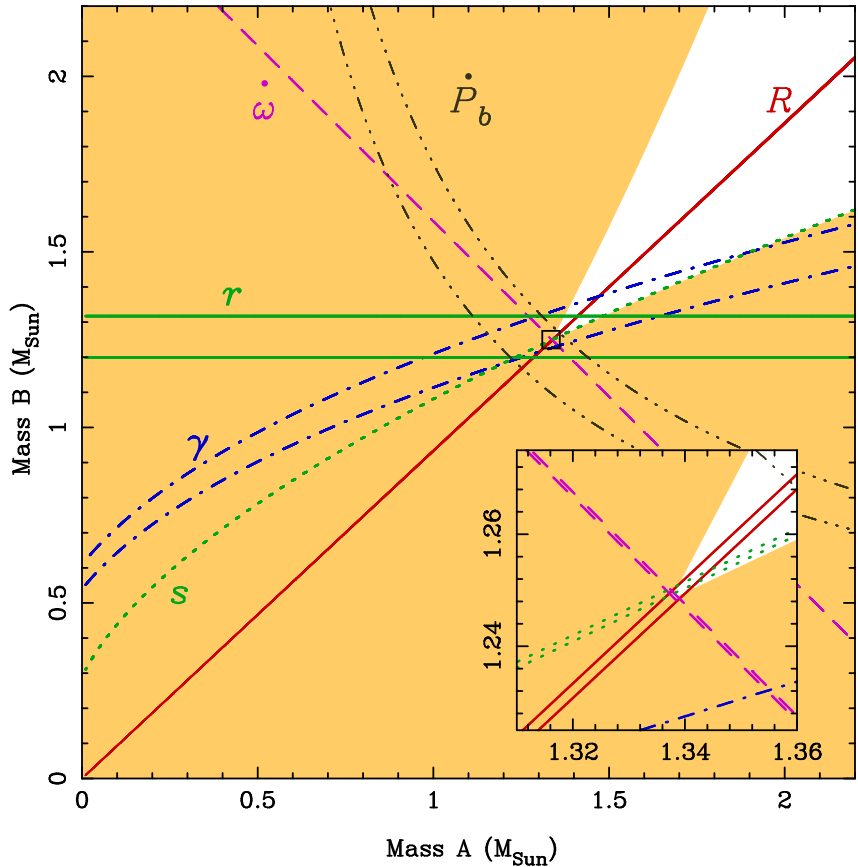}
\caption{\label{fig:m1m2} `Mass--mass' diagram showing the
observational constraints on the masses of the neutron stars in the
double pulsar system J0737--3039.  The shaded regions are those that
are excluded by the Keplerian mass functions of the two
pulsars. Further constraints are shown as pairs of lines enclosing
permitted regions as given by the observed mass ratio and PK
parameters shown here as predicted by general relativity.
Inset is an enlarged view of
the small square encompassing the intersection of these
constraints (see text).}
\end{figure*}

Our observations already provide measurements for all five PK
parameters listed above. This includes a measurement of an orbital
decay of the binary orbit 
due to gravitational wave emission with a rate of
7mm/day.  As indicated earlier, we can use these results to test GR in
a very elegant way \cite{dt92}.  The unique relationship between the
two masses of the system predicted by GR (or any other theory) for
each PK parameter can be plotted in a diagram showing the mass of A on
one axis and B on the other. We expect all curves, including that of
the mass ratio $R$, to intersect in a single point if the chosen
theory (here GR) is a valid description of the nature of this
system. Such tests have been possible to date in PSR B1913+16
(e.g.~\cite{wt03}) and for PSR B1534+12 (e.g.~\cite{sttw02}). However,
in neither of these systems were so many curves available as for the
double pulsar system for which we derive a $M_A-M_B$ plot as shown in
Fig.~\ref{fig:m1m2}.

\begin{table*}
\caption{Observed and derived parameters of PSRs~J0737$-$3039A and B.
Standard errors are given in parentheses after the values and are in
units of the least significant digit(s).}

%\begin{center}
\footnotesize
\begin{tabular}{lcc}
\hline
\hline
 & & \\ 
Pulsar & PSR~J0737$-$3039A & PSR~J0737$-$3039B \\
Pulse period $P$ (ms) & 22.699378556138(2) & 2773.4607474(4) \\
Period derivative $\dot{P}$ & $1.7596(2) \times 10^{-18}$ & $0.88(13)\times 10^{-15}$ \\
Epoch of period (MJD) & \multicolumn{2}{c}{52870.0} \\
Right ascension $\alpha$ (J2000) & 
  \multicolumn{2}{c}{$07^{\rm{h}}37^{\rm{m}}51^{\rm{s}}.24795(2)$ } \\
Declination $\delta$ (J2000) & 
  \multicolumn{2}{c}{$-30^\circ 39' 40''.7247(6)$ } \\
Orbital period $P_{\rm{b}}$ (day) & 
  \multicolumn{2}{c}{0.1022515628(2)} \\
Eccentricity $e$ & \multicolumn{2}{c}{0.087778(2) } \\
Epoch of periastron T$_0$ (MJD) & \multicolumn{2}{c}{52870.0120588(3) }\\
Advance of periastron $\dot{\omega}$ (deg yr$^{-1}$) & 
  \multicolumn{2}{c}{16.900(2)}\\ 
Longitude of periastron $\omega$ (deg) & 73.805(1) & 73.805 + 180.0 \\
Projected semi-major axis $x=a{\rm{sin}}i/c$ (sec) & 1.415032(2) & 1.513(4) \\
Gravitational redshift parameter $\gamma$ (ms) & 0.39(2) &  \\
Shapiro delay parameter $s=\sin i$ & $0.9995(4)$ & \\
Shapiro delay parameter $r$ ($\mu$s) & $6.2(6)$ &  \\
Orbital decay $\dot{P}_b$ ($10^{-12}$) & $-1.20(8)$ \\
Mass ratio $R=M_A/M_B$ & \multicolumn{2}{c}{1.071(1) } \\
%Geodetic precession rate (deg yr$^{-1}$) & 4.79 & 5.07 \\
 & & \\
\hline
\end{tabular}
%\end{center}
\end{table*}

It turns out that, as another stroke of luck, we are observing the
system almost completely edge-on which allows us to determine a
Shapiro delay to very high precision. It also means, however, that we
can probe a pulsar magnetosphere for the very first time using a
background beacon. Results of the magnetospheric interactions between
A and B \cite{mkl+04} and the eclipse of A's signal at its superior
conjunction \cite{lbk+04,krb+04,mll+04} have been presented elsewhere
and must be considered in search for a possible contamination of the
timing data and hence a violation of our assumption that we deal
with a ``clean'' system of point sources.

While all studies so far confirm the cleanness of the system, we also
have to consider that the times-of-arrival (TOAs) are obtained with a
standard ``template matching'' procedure that involves a
cross-correlation of the observed pulse profile with high
signal-to-noise ratio template (e.g.~\cite{tay92}). Any change in the
pulse profile could therefore lead to systematic variations in the
measured TOAs. For this reason it was necessary to perform detailed
studies of the profiles of A and B and to investigate any possible
profile evolution with time. Indeed, profile changes on secular time
scales are expected for the double pulsar for the following reason.

In GR, the proper reference frame of a freely falling object suffers a
precession with respect to a distant observer, called geodetic
precession. In a binary pulsar system this geodetic precession leads
to a relativistic spin-orbit coupling, analogous of spin-orbit
coupling in atomic physics \cite{dr74}.  As a consequence, the pulsar
spins precess about the total angular momentum, changing the relative
orientation of the pulsars to one another and toward Earth. Since the
orbital angular momentum is much larger than the pulsars' angular
momenta, the total angular momentum is practically represented by the
orbital angular momentum. 
The precession rate \cite{bo75} depends on the period
and the eccentricity of the orbit as well as the masses of A and B.
With the orbital parameters of the double pulsar, GR predicts
precession periods of only 75~yr for A and 71~yr for B.

Geodetic precession should have a direct effect on the timing as it
causes the polar angles of the spins and hence the effects of
aberration to change with time \cite{dt92}.  These changes modify the
{\em observed} orbital parameters, like projected semi-major axis and
eccentricity, which differ from the {\em intrinsic} values by an
aberration dependent term, potentially allowing us to infer the system
geometry \cite{sta04c}.  
Other consequences of geodetic precession can be expected
to be detected much sooner and are directly relevant for the timing of
A and B. These arise from variations in the pulse shape due to
changing cuts through the emission beam as the pulsar spin axes
precess.  Moreover, geodetic precession also leads to a change in the
relative alignment of the pulsar magnetospheres, so that the
visibility pattern or even the profile of B may vary due to these
changes as well.

The possibility to observe such phenomena is exciting but requires a
careful analysis to exclude any impact onto the timing of the two
pulsars. Indeed, the necessity to check for these effects has delayed
the publication of the currently final timing results of the double
pulsar. The analysis is now almost complete and results will be published
shortly. A preliminary update is given in Table 1 and discussed below.

\section{PRESENT RESULTS}

The study of the profile evolution of A \cite{mkp+05} did not lead to
the detection of any profile change over a period of 15 months. This
present non-detection greatly simplifies the timing of A but does not
exclude the possibility that changes may not happen in the
future. While the effects of geodetic precession could be small due to
a near alignment of pulsar A's spin and the orbital momentum vector,
the results could also be explained by observing the system at a
particular precession phase. While this case appears to be relatively
unlikely, it must not be excluded as such a situation had indeed
occurred for PSR B1913+16 \cite{kra98}.  Indeed, a modelling of the
results suggests that this present non-detection of profile changes is
consistent with a rather wide range of possible system geometries. One
conclusion that can be drawn, however, is that the observations are
inconsistent with the large profile changes that had been predicted by
some models \cite{jr04}.

In contrast to the results for A, similar studies of the profile
evolution of B \cite{bpm+05} reveal a clear evolution of B's emission
on orbital and secular time-scales. The profile of B is
changing with time, while also the light-curves of B (i.e.~the
visibility of B versus orbital phase) undergo clear changes.  These
phenomena may be caused by a changing magnetospheric interaction due
to geometry variations resulting from geodetic precession. In any case,
these changes require sophisticated timing analysis techniques and the
preliminary results obtained with this techniques are listed in Table
1, while final results will be published shortly.

The present timing results already indicate that the proper motion of
this system is surprisingly small. While a significant measurement of
a proper motion vector via pulsar timing will be available shortly,
the present limit suggests a systemic velocity of less than 30 km/s
for a dispersion measure distance of 600 pc 
\cite{cl02}.\footnote{We note that this
small velocity is consistent with the 66 km/s derived from
scintillation measurements as the latter value is not corrected for a
relative motion of the Earth \cite{cmr+05}.}  While such a small
velocity may be indicative of a small kick imparted onto B during its
supernova explosion \cite{ps05}, other studies find this limit still
to be consistent with a kick of average magnitude \cite{wk05}. In any
case, such a small velocity is good news for tests of GR. Usually, the
observed value of $\dot{P}_b$ is positively biased by an effect known
as secular acceleration arising from a relative motion and
acceleration of the system (e.g.~\cite{dt91}). Computing the magnitude
of this observational bias using the obtained limit on the proper
motion, however, suggests that the contribution is much less than 1\%,
so that the orbital decay measurement will be available for another
precise GR test.

\begin{figure}
\includegraphics[width=70mm]{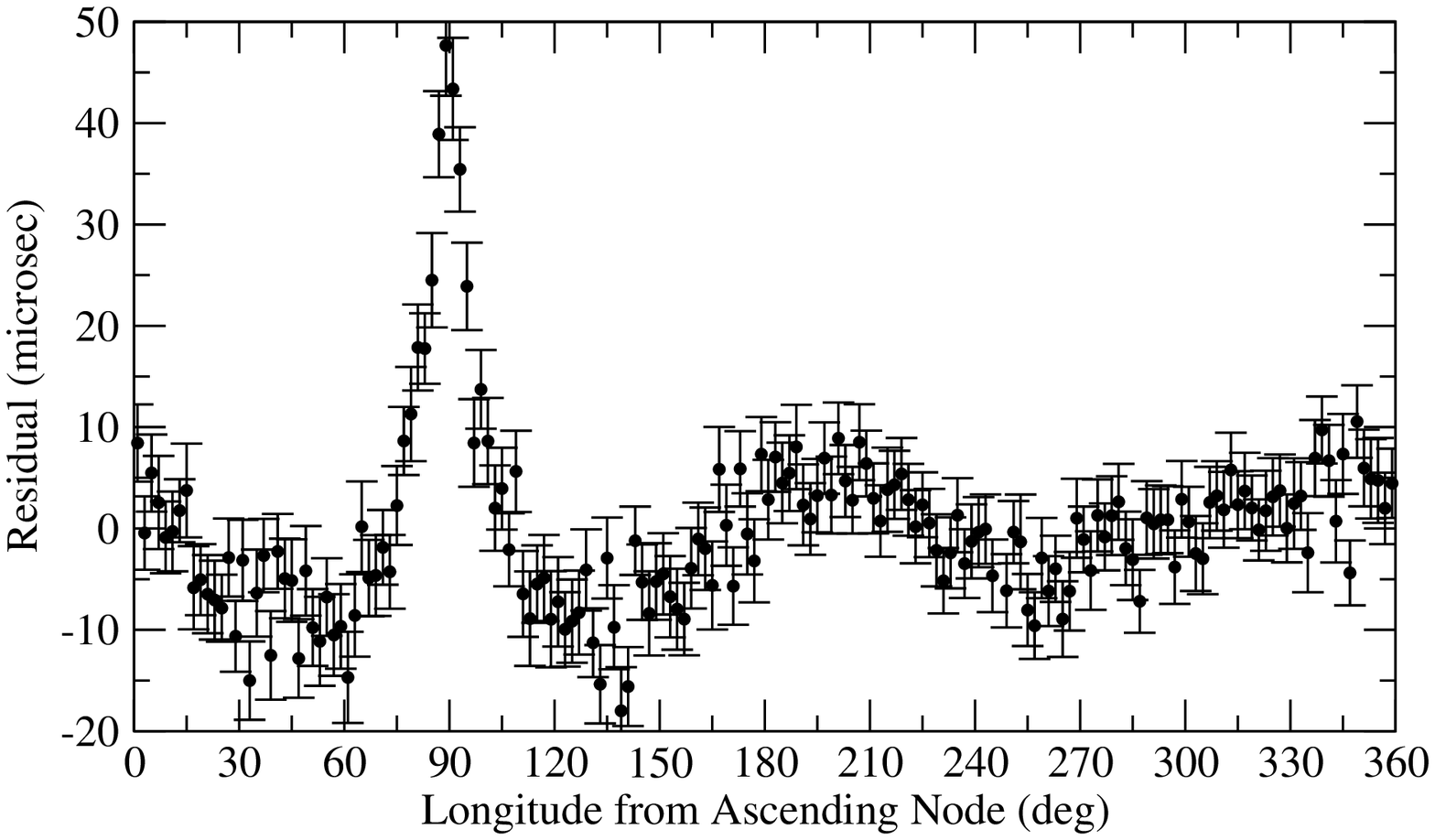}
\caption{ The effect of the Shapiro delay caused by the gravitational
potential of B seen in the timing residuals of A. The timing residuals
obtained by fitting all model parameters shown in Table 1 except the
Shapiro delay parameters $r$ and $s$.  The left-over structure
represents the higher harmonics of the Shapiro delay that are
unabsorbed by fits to the Keplerian parameters.
\label{fig:shapiro} }
\end{figure}

Finally, scintillation measurements have recently suggested an orbital
inclination angle that is extremely close to $90\deg$ (i.e.~within
$(0.29\pm0.14)\deg$) \cite{cmr+05}.  This measurement appears to be
inconsistent with results from the timing observations and the
measurement of the Shapiro delay parameter, $s=\sin i$, which suggest
an inclination angle that is close but significantly different from
$90\deg$. One should note that the scintillation results are based on
correlating the scintillation properties of A and B over
the short time-span of the orbital motion when
they are in conjunction to the observer. In contrast, the
measurement of the inclination angle from timing measurements results
from detecting significant harmonic structure in the post-fit
residuals after parts of the Shapiro delay effect are absorbed in the
fit for the R{\"o}mer delay, i.e.~the light travel time across the
orbit.  As shown in Figure~\ref{fig:shapiro}, these structures are
present throughout the whole orbit, so that the results from
timing measurements may be expected to be more reliable.  We are
currently studying the origin of this apparent inconsistency between
these two methods, checking both any contamination of the Shapiro
delay measurements and effects influencing the scintillation
results. An exciting possibility could be that the
emission of A suffers measurable refraction while propagating through
the magnetosphere of B. If that were indeed the case, we would have a
direct handle onto the magneto-ionic properties of B's magnetosphere
for the first time.

Inspecting the results shown in Table~1, we can take the most
precise parameters (i.e.~the mass ratio $R$, the advance of periastron
$\dot{\omega}$ and the Shapiro delay parameter $s$) to test 
theories of gravity. Assuming that GR is the correct theory
of gravitation, we use Eqn.~\ref{omegadot} to derive the total
mass of the system and combine it with the observed mass ratio
to obtain $M_A=1.338\pm0.001M_\odot$ and $M_B=1.249\pm0.001
M_\odot$. Using these precisely determined masses we compute
the Shapiro delay parameter $s$ as predicted by GR and compare
it to the observed value. We find that $s^{\rm GR}/s^{\rm obs}
= 1.0002^{+0.0011}_{-0.0006}$. Hence, GR passes this test at
the $0.1$\% level. This is the most stringent test of GR in
the strong-field limit so far.

\section{FUTURE}

In the near and far future, the precision of the determined parameters
will increase further, simply by the available longer time span but
also by the potential employment of better instrumentation. In a few
years, we should be able to measure additional PK parameters,
including those which arise from a relativistic deformation of the
pulsar orbit (resulting in angular and radial orbital
eccentricities) and those which find their origin in aberration
effects and their interplay with geodetic precession (see
\cite{dt92}).  On secular time scales we will even achieve a precision
that will require us to consider post-Newtonian terms that go
beyond the currently used description of the PK parameters. Indeed,
the equations for the PK parameters given earlier are only correct to
lowest PN order. However, higher-order corrections are
expected to become important if timing precision is sufficiently high.
While this has not been the case in the past, the double pulsar
system may allow measurements of these effects in the future
\cite{lbk+04}.

One such effect involves the prediction by GR that, in contrast to
Newtonian physics, the neutron stars' spins affect their orbital
motion via spin-orbit coupling. This effect would be visible most clearly
as a contribution to the observed $\dot{\omega}$ in a secular
\cite{bo75} and periodic fashion \cite{wex95}.  For the J0737$-$3039
system, the expected contribution is about an order of magnitude
larger than for PSR B1913+16, i.e.~$2\times 10^{-4}$ deg yr$^{-1}$
(for A, assuming a geometry as determined for PSR B1913+16
\cite{kra98}). As the exact value depends on the pulsars' moment of
inertia, a potential measurement of this effect allows the moment of
inertia of a neutron star to be determined for the first time
\cite{ds88}.

If two parameters, e.g.~the Shapiro parameter $s$ and the mass ratio
$R$, can be measured sufficiently accurate, an expected
$\dot{\omega}_{\rm exp}$ can be computed from the intersection
point. This value can be compared to the observed value
$\dot{\omega}_{\rm obs}$ which is given by (see \cite{ds88})
\begin{equation}
\dot{\omega}_{\rm obs} = \dot{\omega}_{\rm 1PN}\;\left[
1+ \Delta\dot{\omega}_{\rm 2PN} 
  - g^A \Delta \dot{\omega}_{\rm SO}^A
  - g^B \Delta \dot{\omega}_{\rm SO}^B
\right]
\end{equation}
where the last two terms represent contributions from the pulsar spin.
In these terms, $g^{A,B}$ are geometry dependent factors whilst
$\Delta \dot{\omega}_{\rm SO}^{A,B}$ arise from relativistic
spin-orbit coupling, formally at the 1PN level.
However, it turns out
that for binary pulsars these effects have a magnitude equivalent to
2PN effects \cite{wex95}, so that they only need to be
considered if $\dot{\omega}$ is to be studied at this higher level of
approximation. We find $\Delta \dot{\omega}_{\rm SO}\propto I/P M^2$
\cite{ds88}, so that with precise masses $M$
the moment of inertia $I$ can be measured and the neutron star
``equation-of-state'' and our understanding of matter at extreme
pressure and densities can be tested.

The dependence of $\Delta \dot{\omega}_{\rm SO}$ on the spin period
$P$ suggests that only a measurement for pulsar $A$ can be
obtained. It also requires that at least two other parameters can
be measured to a similar accuracy as $\dot{\omega}$. Despite being a
tough challenge, e.g.~due to the expected profile variation caused by
geodetic precession, the prospects are promising. Simulations indicate
that a few years of high precision timing are sufficient.

\section{SUMMARY \& CONCLUSIONS}

With the measurement of five PK parameters and the unique information
about the mass ratio, the PSR J0737$-$3039 system provides a truly
unique test-bed for relativistic theories of gravity. So far, GR
also passes this test with flying colors. The precision of
this test and the nature of the resulting constraints go beyond what
has been possible with other systems in the past. The test achieved so
far is, however, only the beginning of a study of relativistic
phenomena that can be investigated in great detail in this wonderful
cosmic laboratory.

%\begin{figure*}[t]
%\centering
%\includegraphics[width=135mm]{JACpic2.eps}
%\caption{Example of full width figure.} \label{JACpic2-f1}
%\end{figure*}

\begin{acknowledgments}
MK is grateful for the local support provided by the conference
organisers.
\end{acknowledgments}

\bigskip % extra skip inserted

%\bibliographystyle{mkbib}
%\bibliography{journals,modrefs,psrrefs,crossrefs}

\end{document}